# Use of waveguides for polarized neutron studies at the micrometric scale

**S.V. Kozhevnikov,**[a,b,*] **T. Keller,**[c,d] **Yu.N. Khaydukov,**[c,d] **F. Ott,**[e,f] **A. Rühm,**[b] **A. Thiaville,**[g] **J. Torrejón,**[g] **M. Vázquez**[h] **and J. Major**[b]

[a] *Frank Laboratory of Neutron Physics, Joint Institute for Nuclear Research,*
    *141980 Dubna, Russian Federation*
[b] *Max-Planck-Institut für Intelligente Systeme (formerly Max-Planck-Institut für Metallforschung),*
    *Heisenbergstr. 3, D-70569 Stuttgart, Germany*
[c] *Max-Planck-Institut für Festkörperforschung,*
    *D-70569 Stuttgart, Germany*
[d] *Forschungsneutronenquelle Heinz Maier-Leibnitz (FRM-II),*
    *D-85748 Garching, Germany*
[e] *CEA, IRAMIS, Laboratoire Léon Brillouin,*
    *F-91191 Gif sur Yvette, France*
[f] *CNRS, IRAMIS, Laboratoire Léon Brillouin,*
    *F-91191 Gif sur Yvette, France*
[g] *Laboratoire de Physique des Solides, Université Paris-sud, CNRS, UMR 8502,*
    *F-91405 Orsay, France*
[h] *Instituto de Ciencia Materiales, CSIC,*
    *28049 Madrid, Spain*
    *E-mail:* kozhevn@nf.jinr.ru

ABSTRACT: We present the use of planar waveguides to produce neutron microbeams in order to investigate magnetic micro-structures with a micrometric spatial resolution. We report experimental results on such measurements on a polarized neutron reflectometer using a nonmagnetic waveguide structure to probe an amorphous magnetic microwire. The spin-resolved transmission of a neutron microbeam through a microwire was studied. The one dimensional mapping of the neutron precession across the wire was measured.

---

[*] Corresponding author.

**Contents**

## 1. Introduction

Progress in nanotechnologies requires new methods for nanostructures characterization. Neutron scattering is a powerful method for the investigation of biological objects, polymers and magnetism but the spatial resolution is rather limited. To reduce the beam size, focusing devices (bent crystals, zone plates, compound refractive lenses, etc.) are being developed [1]. However, these devices have limitations which prevent obtaining beam sizes smaller than 50μm. Other possible focusing devices are planar waveguides which can transform an initial macrobeam into a very narrow, though slightly divergent microbeam. This idea was successfully realized for x-rays [2-4]. In the case of neutrons, the practical realization is more complicated. The reason is the difference between the properties of x-rays and neutron beams. A synchrotron source produces a highly collimated beam of high intensity. Also, the x-rays have a strong interaction with matter and can be efficiently reflected from surfaces. On the other hand, neutron sources produce very divergent beams of relatively low intensity. Neutron interaction with matter is weak and therefore it is rather difficult to focus neutron beams. Early experiments using waveguides with a *prism-like coupler* demonstrated the production of unpolarized [5] and polarized [6] neutron microbeams. Later measurements using a *resonant beam coupler* [7] also showed the production of unpolarized neutron microbeams. However, neutrons microbeams were never used for practical measurements until now.

    In our previous works, the production of polarized neutron microbeams from polarizing and non-polarizing magnetic waveguides was demonstrated [8]. The combination of a nonmagnetic planar waveguide and a polarized neutron reflectometer was used for the demonstration of polarization analysis of a neutron microbeam [9]. In the article [10], a conventional polarized neutron beam was used in the regime of spin-resolved transmission for the investigation of domain walls in a magnetic film. In the present communication, we report the first practical application of a polarized neutron microbeam with a width of the order of 2μm

for the investigation of a magnetic microwire using spin-resolved transmission. One of the key issues of such experiments is the very low neutron signal. We thus discuss in details the microbeam extraction from the nonmagnetic waveguide and the background suppression which led to the feasibility of such measurements.

## 2. Waveguides for the production of microbeams

### 2.1 Principle of planar waveguides

A typical waveguide structure comprises of a low optical potential layer sandwiched between two cladding layers (figure 1). When a neutron beam is incident on such a structure, the neutrons tunnel into the guiding layer through the thin upper layer (coupling layer) and are reflected from the bottom reflecting layer of high optical potential. The neutrons are guided inside the channel over the channeling length which is of the order of a few mm [11] and eventually exit the structure in the specular direction. However, the neutrons incident on the very last part of the sample are guided down to the edge of the channel and eventually exit the edge of the waveguide channel. These neutrons thus create a very localized source of neutrons with a typical size of the order of 100 nm. The neutrons exiting the channel edge are however subject to Fraunhofer diffraction which gives rise to a significant divergence. For a typical sample, it is of the order of 0.15°. The width of the microbeam exiting the sample edge thus quickly expands but after a travel of 1 mm in the air the beam size is still as small as 2.6 µm. The shape of the microbeam is thus a very narrow vertical slit 2 µm wide × 20 mm high (substrate size).

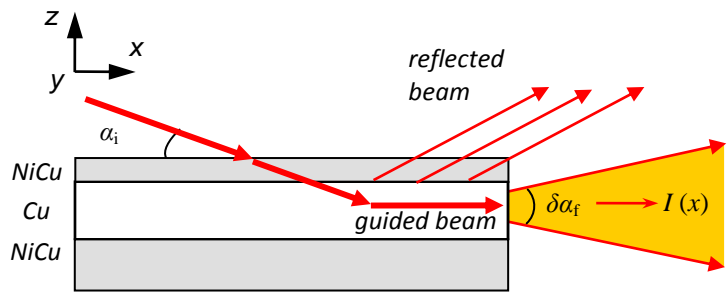

**Figure 1.** Sketch of a waveguide structure for the production of a neutron microbeam. An incident neutron beam tunnels into the guiding layer through the thin upper layer (coupling layer), is guided inside the channel and eventually exits the structure either in the specular direction or by the sample edge thus producing a microbeam.

Figure 2 describes the wavefunction properties in such a wave-guide structure. We have considered the structure $Ni_{0.67}Cu_{0.33}(20nm)/Cu(150nm)/Ni_{0.67}Cu_{0.33}(50nm)//Si(substrate)$ (figure 2a). The theory of resonances in layered structures can be found in [12]. The wavefunction density is shown in figure 2b as a function of the incidence $\alpha_i$ and the perpendicular coordinate $z$ below the waveguide surface. Inside the guiding layer (or channel), the neutron wavefunction density is resonantly enhanced and has density maxima up to 30 (normalized to the incident wavefunction density). The calculations were made with the program *SimulReflec* [13]. One can see one, two or three maxima in the direction perpendicular to the waveguide surface and in the region of angles below the critical angle for total reflection. These maxima are marked by the indices 0, 1, 2 and 3 corresponding to the resonance orders $n$=0, 1, 2, 3.

The wavefunction density at the resonance angles $\alpha_{i0} = 0.358°$, $\alpha_{i1} = 0.375°$, $\alpha_{i2} = 0.400°$ is shown in figure 2c as a function of the coordinate $z$. One can see that the wavefunction density at the resonance $n$=0 has a maximum in the centre of the guiding layer. The wavefunction density at the resonance $n$=1 has two maxima close to the interfaces of the guiding layers and has a minimum in the centre. At the resonance $n$=2, the wavefunction density has three maxima in the centre and close to the interfaces.

In figure 2d, the total wavefunction density (integrated over the coordinate $z$ inside the guiding layer) is presented as a function of the initial angle $\alpha_i$. One can see the maxima at the resonance angles $\alpha_{in}$. The amplitude of the maxima decreases with increasing resonance order.

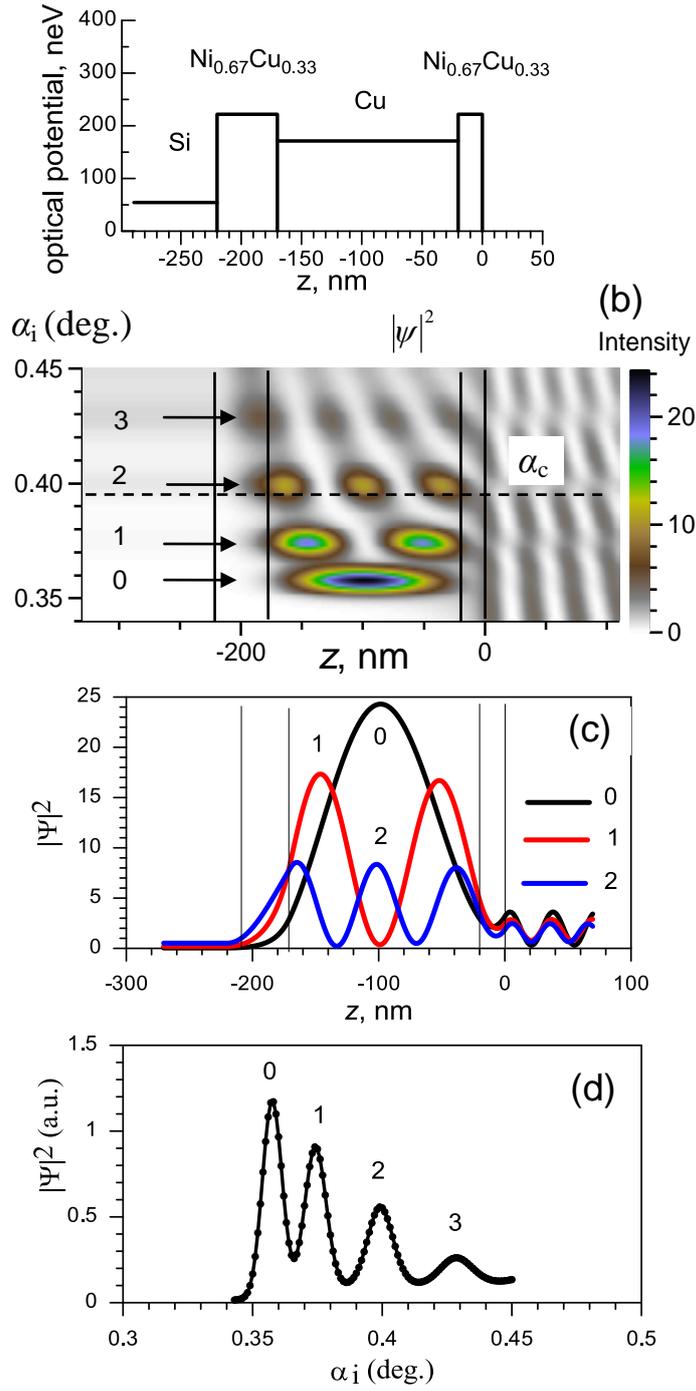

**Figure 2.** Calculations for the nonmagnetic waveguide structure using the program *SimulReflec* [13]. (a) Neutron optical potential as a function of the depth $z$ inside the waveguide. (b) Neutron wavefunction density as a function of the incidence angle $\alpha_i$ and the depth $z$. (c) Neutron wavefunction density as function of the depth $z$ at the incident angles corresponding to the resonances $\alpha_{i0} = 0.358°$; $\alpha_{i1} = 0.375°$; $\alpha_{i2} = 0.400°$. (d) Neutron wavefunction density integrated over the depth $z$ inside the guiding layer as a function of the incidence angles $\alpha_i$.

## 2.2 Specular reflectivity and background suppression

Characterization experiments were performed on the horizontal sample plane reflectometer NREX (at the research reactor FRM II, Garching, Germany) in unpolarized mode. The neutron wavelength was 4.26 Å (2 % wavelength resolution) and the incidence angular resolution was 0.012°. The two-dimensional $^3$He PSD had a 2.5 mm spatial resolution. The distance 'first slit - waveguide' was 2200 mm and the distance 'waveguide - detector' was 2500 mm.

In order to be able to observe the microbeam, it is necessary to reduce parasitic signals and in particular to mask the reflected beam which is several orders of magnitude more intense than the micro-beam. This was achieved by using an absorbing bar (borated aluminium, boral) of size 35×1×1 mm$^3$ which was placed on the waveguide surface close to the exit edge (figure 3a). Note that the neutron channeling length, that is the length over which neutrons are carried inside the guide, is 1.7 mm for this system. Thus, in order to prevent losses of guided neutrons, the length of the absorber at the sample edge should thus be as short as possible hence the choice of 1 mm.

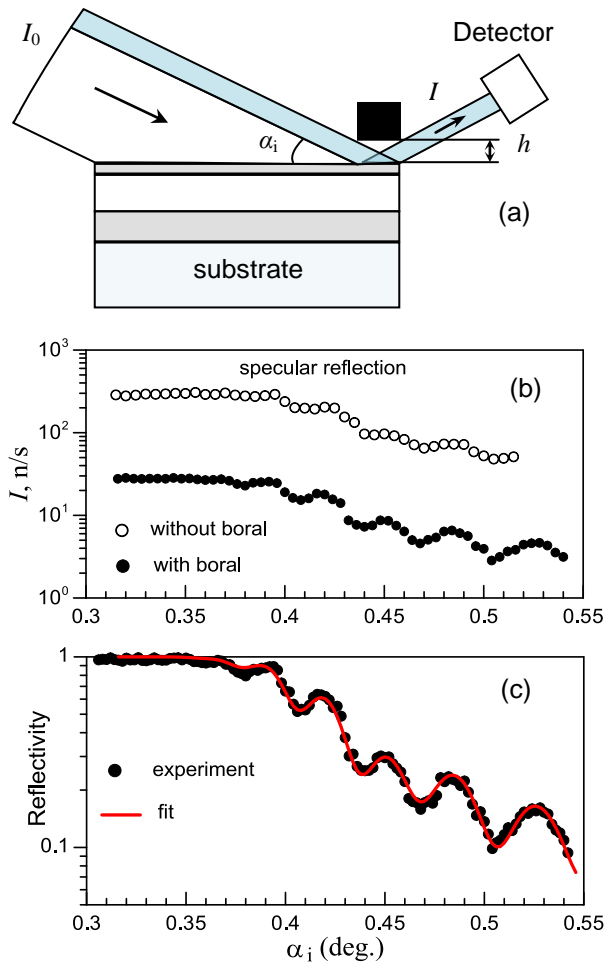

**Figure 3.** (a) Scheme of the measurements with an absorbing bar on the waveguide horizontal surface (in practice there is an air gap $h$ between the 2 elements): $D$ is the width of the absorbing bar. (b) Specular reflection intensity without (open symbols) and with the boral absorber (closed symbols). (c) Fit (solid line) of the specular reflectivity with the absorbing bar.

The absorbing bar was treated by electro-polishing. However defects during the preparation led to an air gap $h$ between the surfaces of the waveguide and the absorber. In figure 3b, the specularly reflected beam intensity without any absorber (open symbols) and with the absorber is shown. The specular reflected beam intensity with the absorber (30 n/s) is about 10% of the intensity without the absorber (300 n/s). The air gap transmits only 10% of the specular reflected intensity. The gap height can thus be estimated as $h \approx 11$ μm.

The measured and fitted specular reflectivity with the absorbing boral bar is shown in figure 3c. The following structure was deduced from the fit: CuO(2.5 nm)/NiCu(14.9 nm)/Cu(141.7 nm)/NiCu(53.3 nm)//Si(substrate). The following nuclear potentials for the different layers were found: CuO (45 neV), upper layer NiCu (245 neV), Cu(171 neV), lower layer NiCu (219 neV), Si (54 neV). The known nuclear potential of the $Ni_{0.67}Cu_{0.33}$ alloy is 229 neV. The fitted parameters are close to this value.

## 2.3 Microbeam mapping

A two-dimensional intensity map $I(\alpha_i, \alpha_f)$ is shown in figure 4a. The boral absorber was in place on the waveguide surface. The upper diagonal corresponds to the specular reflected beam and the bottom diagonal to the direct and transmitted beams. The microbeams can be observed around the horizon $\alpha_f = 0$ and are marked by the indices of the resonance number $n$=0, 1 and 2 and by vertical lines at the incidence angles $\alpha_{in}$. The spots of the microbeams maxima are marked by circles. Thanks to the boral absorber, the background is low enough so that even the resonances of higher orders $n$=1 and 2 can be observed in the experiment. One can see that the two-dimensional intensity map in figure 4a reproduces the wave function density distribution $|\psi(z,\alpha_i)|^2$ inside the guiding layer (figure 2b). Note however that figure 2b is the wave function density inside the layer while the map of figure 4a represents the beam intensity after diffraction from the edge of the guiding layer.

In figure 4b, the microbeam intensity integrated between the dotted lines (figure 4a) is shown as a function of the incidence angle $\alpha_i$. The triangle and circle symbols correspond to the intensity without and with the boral absorber, respectively. For the resonance $n$=0, the microbeam intensity is about 8 n/s. The ratio *signal/background* is about 8/3=2.7 and 8.5/1.5=5.7 for the cases without and with the boral absorber, respectively. Thus, the use of the boral absorber increases the ratio *signal/background* by a factor 2. The dependence of the microbeam intensity as a function of the resonance order follows the dependence of the neutron wave function density calculated in figure 2d. Note that the presence of the boral absorber only reduces the background from the reflected beam while the microbeam intensity is unaffected (figure 4b).

In figure 4c, the microbeam intensity at the resonance incidence angles $\alpha_{i0} = 0.360°$; $\alpha_{i1} = 0.378°$ and $\alpha_{i2} = 0.400°$ is shown as a function of the final angle $\alpha_f$. The microbeam intensity distribution is symmetric with respect to the horizon $\alpha_f = 0$. At the right side around $\alpha_f = 0.3°$ the part of the specular reflected beam is shown. One can see that the intensity distribution has one, two and three maxima for the resonances $n$=0, 1 and 2, respectively. This is the consequence of the $z$ coordinate dependence of the wave function density inside the guiding layer (figure 2c). The microbeam of the resonance $n$=0 has a one-peak Gaussian distribution and the highest intensity (figure 4c). This mode is the most convenient for the applications to microstructure investigations. The experimental angular divergence of this microbeam is $\delta\alpha_f = 0.2°$ (FWHM).

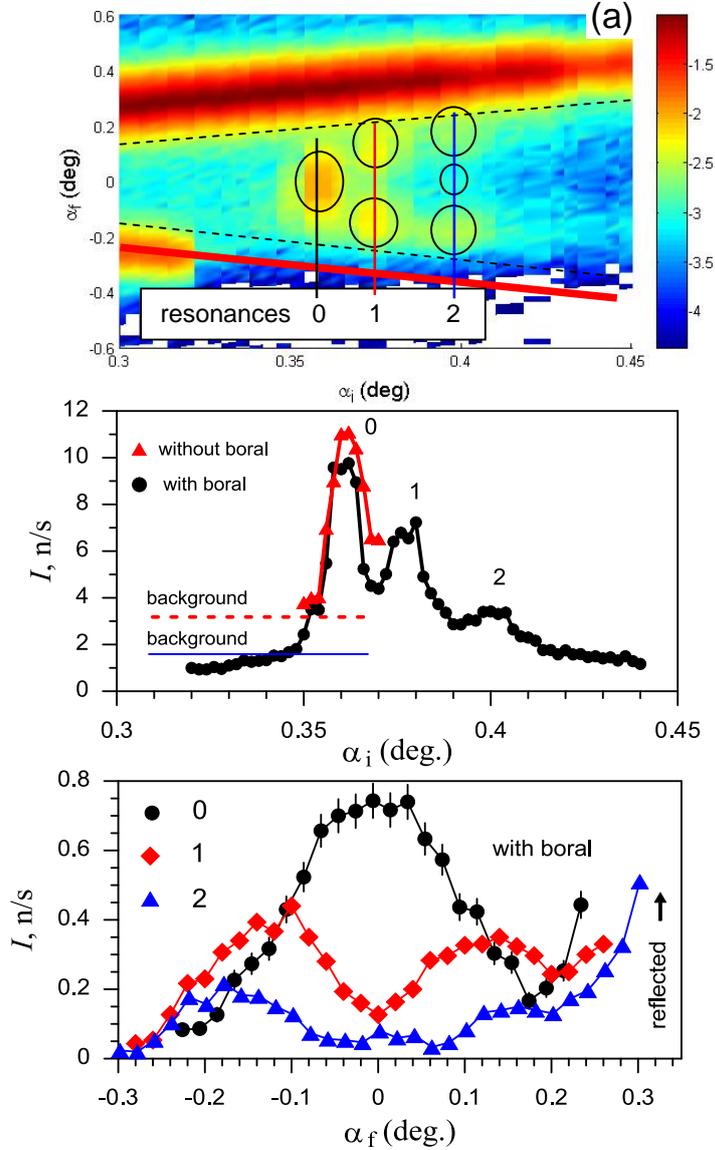

**Figure 4.** Mapping of the microbeam intensity produced by a nonmagnetic waveguide with an absorber on the surface. (a) Two dimensional intensity map $(\alpha_i, \alpha_f)$: the upper diagonal is the specular reflected beam; the bottom diagonal is the direct beam cut by the detector window at the angle $\alpha_{i0} = 0.32°$; the horizon is defined by $\alpha_f = 0$; the indices 0, 1 and 2 correspond to the resonance orders. (b) Microbeam intensity integrated between the dashed lines with and without the boral absorber. (c) Microbeam intensity distribution of the different resonance orders as a function of the exit angle $\alpha_f$.

## 3. Use of polarized microbeams for magnetic microstructure investigations

### 3.1 Experimental setup

The experiment was performed on the vertical film surface polarized neutron reflectometer PRISM [14] at the reactor Orphée (LLB, Saclay, France). The scheme of the experiment (top view) is shown in figure 5. The monochromatic incident neutron beam is transmitted through a supermirror polarizer (not shown), collimated by a first Cd slit (not shown) and passes the first spin-flipper of the Mezei type (not shown). The neutron wavelength is $\lambda=4.0$ Å (7 % wavelength resolution) and the angular divergence of the beam is 0.02°. The polarized neutron beam of intensity $I_0$ enters the waveguide (WG) surface under the grazing incidence angle $\alpha_i$.

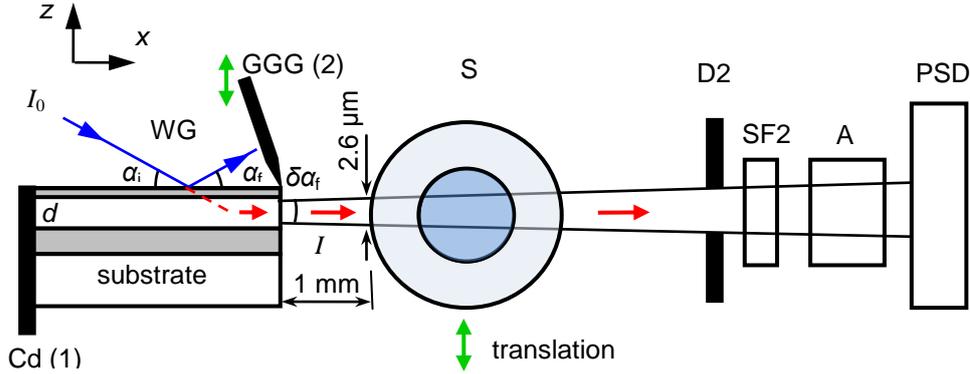

**Figure 5.** Top view of the experimental setup with a wave-guide set vertically. The polarizer, the first slit and the first spin-flipper are not shown. WG is the planar waveguide; Cd is the absorbing plate; GGG is the absorbing crystal blade; S is the microwire; D2 is the second slit; SF2 is the second spin-flipper; A is the analyzer and PSD is the two-dimensional position-sensitive detector.

In order to reduce the background noise, the direct beam is blocked by a Cd absorbing plate (1) glued on the front edge of the waveguide. The reflected beam is stopped by a single crystal GGG (Gadolinium Gallium Garnet) blade (2) which can be adjusted to lay on the surface by a micrometer translation unit. The width of the guiding layer being $d=150$ nm, the size of microbeam is about 2.6 µm at a distance of 1 mm after the wave-guide structure.

The polarized microbeam is then transmitted through the investigated magnetic wire (S) of diameter 190 µm which is set as close as possible to the waveguide exit (1 mm in our experiment). The microwire is fixed vertically in a slightly strained state onto precise translation and tilt stages. The translation allows scanning the wire position across the microbeam. The goniometer is required for the adjustment of the wire to be perfectly parallel with respect to the plane of the neutron guide. The parallelism of the microbeam and the wire can be checked optically (by eye) by aligning the waveguide edge with the wire at grazing incidence. It can also be checked by scanning the tilt angle and measuring the transmission. Since the microwires contain boron $(Co_{0.94}Fe_{0.06})_{72.5}Si_{12.5}B_{15}$, they measurably absorb neutrons (~50%). Thus the absorption maximum corresponds to a perfectly aligned wire. Note that since the microbeam and the reflected beam are perfectly parallel, such an alignment can be performed also with the reflected beam (by removing the GGG beam trap).

A second slit D2 (0.5-1.5 mm) set between the wire and the detector allows to further reduce the background. The neutron beam passes a second Mezei spin-flipper, is transmitted through a supermirror analyzer (A) and is recorded by a two-dimensional $^3$He position-sensitive detector (PSD). The distance 'slit D1 - waveguide' is 1900 mm, the distance 'waveguide - slit D2' is 400 mm and the distance 'waveguide - detector' was 1300 mm. All the parts (waveguide, GGG slit, wire, translations and goniometer) are assembled on a single rotating platform which also supports a set of three Helmholtz coils pairs which allows applying a magnetic field in an arbitrary direction of space.

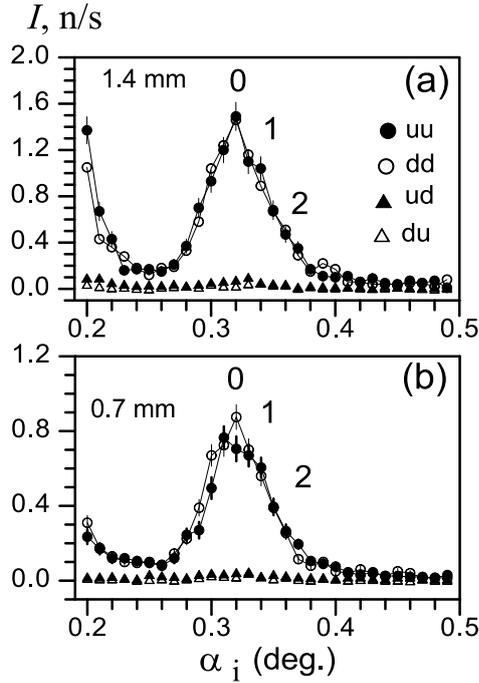

**Figure 6.** The microbeam intensity from the nonmagnetic waveguide for the different polarization modes of the reflectometer as a function of the initial angle $\alpha_i$ for two widths of the slit D2 before the analyzer: (a) 1.4 mm and (b) 0.7 mm. The direct beam is blocked by the Cd plate on the front edge and the reflected beam is suppressed by the GGG absorbing blade near the waveguide surface.

Figure 6 shows a mapping of the microbeam intensity for the different spin states (*up-up*, *down-down*, *up-down*, *down-up*). The detector was fixed at the horizontal position of $\alpha_f = 0$. The indices 0, 1 and 2 mark the positions of the corresponding resonances $n$=0, 1 and 2. Two widths of the diaphragm D2 before the analyzer were used: 1.4 mm (figure 6a) and 0.7 mm (figure 6b). On the left side of the microbeam peak, one can observe the tail from the direct beam passing above the waveguide surface which contributes to the background. For the diaphragm 1.4 mm, the signal is 1.5 n/s and the background is 0.15 n/s. The *signal*/*background* ratio is thus equal to 10. For the diaphragm of 0.7 mm, the signal is 0.8 n/s and the background is 0.05 n/s. The *signal*/*background* is thus equal to 16. The narrow diaphragm D2 reduces the microbeam intensity by a factor of 2 but only improves the *signal*/*background* ratio by a factor of 1.6.

For the empty beam without a waveguide, the flipping ratio (non spin-flip intensity divided by the spin-flip) is 35. For the microbeam, the raw flipping ratios remain close to these values. It demonstrates that the nonmagnetic waveguide does not change the polarization of the incident polarized beam and that the background level is very low.

**3.2 Precession mapping on magnetic microwires**

It was recently demonstrated that neutron precession could be used to probe the inner structure of microscopic objects [10]. In this section we present experimental data where the neutron precession through a magnetic microwire was measured using a polarized microbeam. In such measurements all neutrons are used which makes it possible to work with very low incident neutron fluxes.

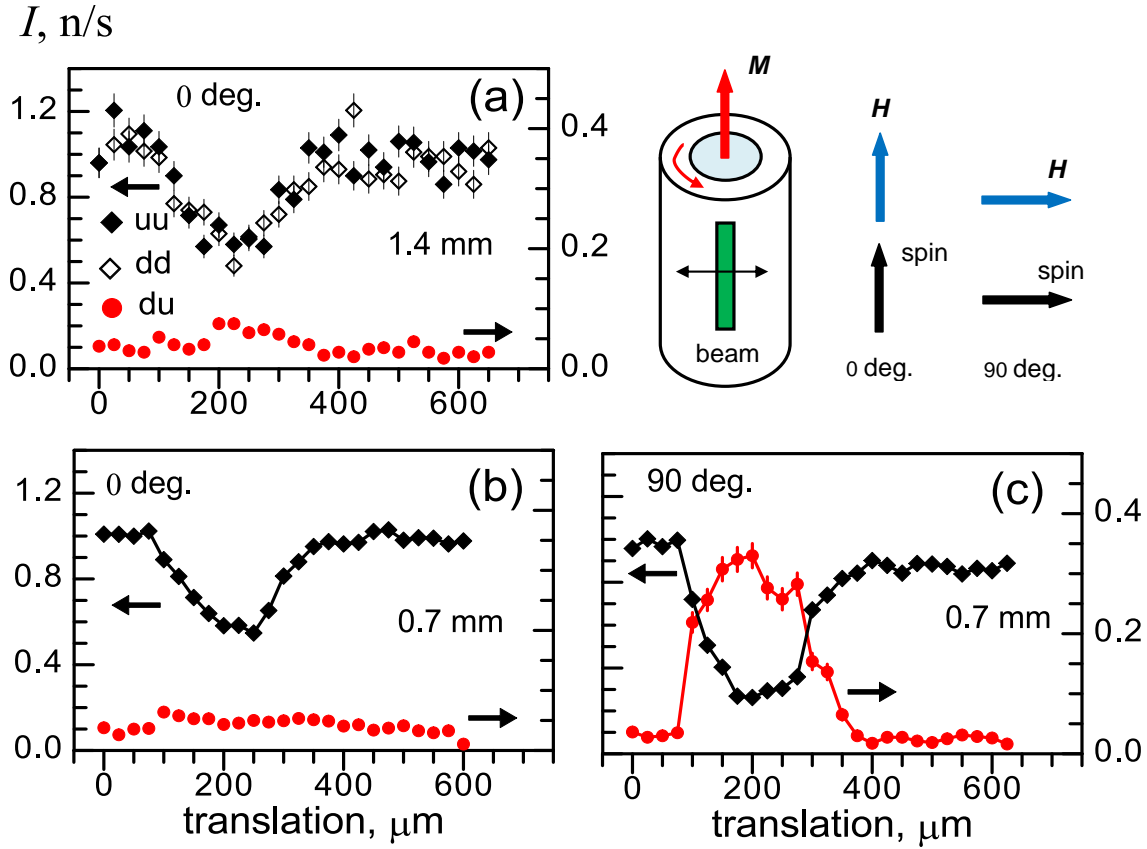

**Figure 7.** Experimental results of the spin-resolved transmission of the polarized microbeam through the magnetic microwire. An external magnetic field of 20 Oe was applied parallel (0 deg.) or perpendicular (90 deg.) to the wire axis. The microbeam intensity for different polarization modes as a function of the wire position: (a) parallel external field (D2=1.4 mm); (b) parallel external field (D2= 0.7 mm); (c) perpendicular external field (D2=0.7 mm).

The internal structure of amorphous magnetic microwires is expected to be complex [15]. In particular a core vortex state is expected to be present in the middle of the wire. In the case of non axial applied fields, non collinear magnetic structures are expected to develop across the wire diameter. These magnetic structures are at the limit of the problems which can be tackled by micromagnetic calculations (in term of size). The motivation of this investigation is thus the experimental determination of the internal micromagnetic structure of the wire in order to validate analytical and numerical models.

The wire $(Co_{0.94}Fe_{0.06})_{72.5}Si_{12.5}B_{15}$ was 190 μm in diameter. The width of the used microbeam was estimated as 2.6 μm at the microwire position (1 mm after the waveguide). An external magnetic field was applied parallel (0°) or perpendicular (90°) to the wire axis (see inset in figure 7). The intensity of the transmitted microbeam as a function of the microwire position is shown for a parallel magnetic field of 20 Oe and for a detector slit of 1.4 mm (figure 7a) and 0.7 mm (figure 7b). The left ordinates correspond to the non spin-flip intensity and the right axes correspond to spin-flip intensity. The translation steps of the wire were 25 μm. The measuring time at the second slit opening of 1.4 mm was 200 s/point for non spin-flip and 400 s/point for spin-flip intensity. The measuring time at the second slit opening 0.7 mm was 1800 s/point for both intensities. One can see that the transmitted intensity has a minimum with a FWHM of about 250 μm. This is close to the microwire diameter 190 μm. The enlargement may be accounted for by the fact that the wire and the microbeam are not perfectly parallel and also by the resolution of 25 μm due to the step value. The absorption of neutrons is about 40 %. The non spin-flip intensities *uu* and *dd* are equal to each other. The residual spin-flip intensity is due to the limited polarizing efficiency of the polarizer and the analyzer. The error bars in figure 7b correspond to the symbols size.

When a 20 Oe magnetic field is parallel to the wire, one expects the wire to be magnetically saturated. One can nevertheless observe a measurable increase of the spin-flip signal so that the flipping ratio drops down to 12. This suggests that magnetic domains are still present even at rather large magnetic fields.

In the perpendicular magnetic field configuration (figure 7c) the spin-flip signal becomes large. The measuring time was 800 s/point. The spin-flip probability is 35 %. The non spin-flip transmission drops from 60 % to 25 %, respectively (right ordinate axis). The flipping ratio drops down to 1. This is striking since it suggests that the wire is fully depolarizing the neutron beam. One can also note that the transmission spectrum becomes asymmetric and that a magnetic depolarization signal can be observed outside the wire position (between $x$=300-400 μm). This feature cannot be attributed to an experimental artefact since its symmetry can be switched be reversing the field. This suggests that significant asymmetric stray fields appear outside the wire. We are presently unable to explain these features with the existing models for the magnetic structure of these wires. Micromagnetic simulations are under way to be able to understand both the magnitudes and symmetry of the observed precession effects [16].

## 4. Conclusions

We have described the practical implementation of planar waveguides for the production of micron size neutron beams. A polarized neutron microbeam of 2.6 μm width was produced by a planar waveguide. This type of set-up can be easily implemented on any standard polarized neutron reflectometer. We have described the key requirements to minimize the background noise. One of the key advantages of such a set-up is that the neutron beam is produced off the direct beam direction so that very low background levels can be achieved. This method circumvents the limitation of neutron imaging related to the limited neutron detector resolution.

We have experimentally demonstrated that, in spite of the low microbeam intensity, spin-resolved transmission can be used to map a magnetic microstructure within a reasonable measurement time. This was demonstrated by measuring the precession of a polarized neutron microbeam through an amorphous magnetic microwire of diameter 190 μm set at a distance of 1 mm from the exit of the waveguide. The shape of the microbeam was a vertical slit parallel to the wire axis. The internal magnetic structure of the wire was mapped by scanning the wire position across the fixed microbeam.

This method may be applied for the investigation of one-dimensional magnetic structures such as ripple domains, lithographic gratings or vortices in superconductors.

**Acknowledgments**

This work has been supported by the French project IMAMINE 2010-09T.